# Unraveling the nature of quasi van der Waals Epitaxy of magnetic topological insulators Cr: $(Bi_xSb_{1-x})_2Te_3$ on a GaAs (111) substrate through coherently strained interface

Yuxing Ren [1], Lixuan Tai [2], Kaicheng Pan [1], Yueyun Chen [3], Benjamin Z. Gregory [4], Jin Ho Kang [2], Malcolm Jackson [1], Michael Liao [1], Yifei Sun [5], Noah Bodzin [7,8], Kin Wong [7,8], Suchismita Sarker [6], B. C. Regan [3,8], Chee-Wei Wong [2,7], Mark Goorsky [1,7,8], Andrej Singer [4,5,6], Kang L. Wang [1,2,3,7,8]

1. Department of Materials Science and Engineering, University of California, Los Angeles

Los Angeles, CA, 90095

2. School of Electrical and Computer Engineering, University of California, Los Angeles,

Los Angeles, CA, 90095

3. Department of Physics and Astronomy, University of California, Los Angeles,

Los Angeles, CA, 90095

4. Department of Physics, Cornell University, Ithaca, NY, 14853

5. Department of Materials Science and Engineering, Cornell University, Ithaca, NY, 14853

6. Cornell High Energy Synchrotron Source (CHESS), Ithaca, NY, 14853

7. Nanofabrication Laboratory, University of California, Los Angeles, Los Angeles, CA, 90095

8. California Nanosystems Institute, University of California, Los Angeles, Los Angeles, CA, 90095

## Abstract

Quasi van der Waals Epitaxy (qvdWE) has been realized for decades at the interfaces between 3D and 2D materials or van der Waals materials. The growth of magnetic topological insulators (MTI) Cr: $(Bi_xSb_{1-x})_2Te_3$ (CBST) on GaAs (111) substrates for Quantum Anomalous Hall Effect (QAH) is actually one of the examples of qvdWE, which is not well noticed despite the fact that its advantages have been used in growth of various MTI materials. This is distinguished from the growth of MTIs on other substrates. Although the qvdWE mode has been used in many 2D growth on III-V substrates, the specific features and mechanisms are not well demonstrated and summarized yet. Here in this work, we have for the first time shown the features of both coherent interfaces and the existence of strain originating from qvdWE at the same time.

**Main**

Magnetic topological insulator (MTI) like Cr: (Bi$_x$Sb$_{1-x}$)$_2$Te$_3$, or CBST, could achieve quantum anomalous Hall (QAH) effect and spin-orbit torque (SOT) switching in the same structure. This is promising for its future applications in memory or switching applications with its robust surface properties by topological protection. [3] If realized, it can be more energy-efficient than traditional CMOS technologies due to the extremely low stand-by energy consumption. Molecular Beam Epitaxy (MBE) has been demonstrated to be one of the best methods for growing CBST and other MTIs. Among all these different substrates for epitaxial growth (eg. SrTiO$_3$, sapphire, InP, etc), GaAs and some other III-V substrates have stood out due to its smaller lattice mismatch, polarization at interface, etc.

Recently, the specific growth mode of quasi van der Waals epitaxy (qvdWE) has been raised to describe the growth of MTIs on GaAs and some other III-V substrates.[6, 7, 18, 19] The qvdWE mode has been applied to many different chalcogenide growths as well and has been drawing people's attention due to the nature that it can have coherent interface and a good crystallinity.

Unlike Van der Waals Epitaxy which has been seen on many 2D growths, in some van der Waals layers, the van der Waals force as a secondary bonding force can still add some constraints to the epitaxial layer though much weaker than first order bonding. Beginning 1 decade ago the quasi van der Waals epitaxy has already been realized via graphene as an inter-layer.[1,5] Remote epitaxy also utilizes the relatively stronger constraints from the 3D substrate with a 2D interlayer. [2] Similarly, the growth of van der Waals layers on a 3D substrate can also have a strong interaction between epitaxial layer and substrates, [6, 7, 18, 19] as in the case of CBST on GaAs (111). The specific surface potential and ionicity determines the nature of a quasi van der Waals growth, which has the same advantage of van der Waals epitaxy of getting rid of surface roughness while at the same time maintaining a uniform grain orientation as traditional epitaxy.

Despite these reports, there are still no clear conclusions on the root cause of qvdWE and the features of qvdWE have not been well recognized. The typical features are coherent interface and the existence of a strain within a short period of distance. While coherency is much more widely observed for qvdWE, the existence of strain is generally hard to be demonstrated due to the small thickness of strained layers in the van der Waals materials. Our goal in this work is to demonstrate both coherency and the existence of strain at the interface of epitaxial CBST on GaAs (111) substrate. X-ray pole figures and Reciprocal Space Mapping (RSM), in-situ Reflective High Energy Electron Diffraction (RHEED), STEM (Scanning Transmission Electron Microscope) and EDS (Energy Dispersive X-ray Spectroscopy), and a Raman measurement are used to demonstrate the coherently strained interface.

To show the coherently strained interface, we have grown samples with thickness of 1 QL (quintuple layer), 2QL and 6 QL under the same conditions respectively to compare the relaxed layers and strained layers.

Shown in Fig. 1. is a summary of our growth process and some basic structure characterizations.

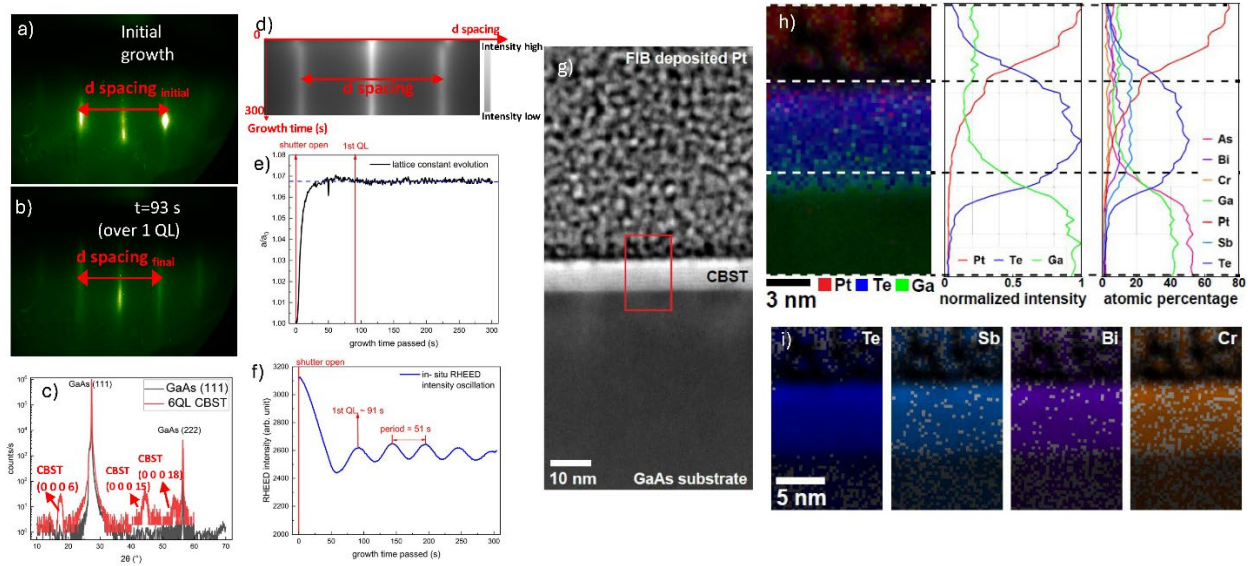

**Fig.1. Growth process of CBST and basic structure characterizations. a)-b)** In-situ Reflective High-Energy Electron Diffraction (RHEED) pattern of GaAs (111) substrate (a)) and CBST (b)). The sample is showing <1 1 -2 0> direction in the RHEED image and is fixed at the same location during growth without rotation. **c)** X-ray diffraction of the grown CBST sample in comparison to bare substrate in ω-2θ scan. It shows clear peaks corresponding to CBST besides the peaks from GaAs substrate. **d)** Recording of d-spacing image during growth. **e)-f)** Time evolution of relative lattice constant value (e)) and RHEED intensity (f)) during growth. The relative lattice constant is compared with initial value of GaAs and is calculated inversely from d-spacing from RHEED image. It shows linear transition at the initial growth from GaAs to CBST and then stabilizes during all the growth. RHEED intensity shows clear oscillation with a period around 51s/QL. It can be seen that the lattice constant relaxes at around 30s which corresponds to the growth of half layer of 1QL. **g)** A high-angle annular dark field (HAADF) image of a 6 QL (quintuple layer) CBST cross section. The Pt layer on top is deposited during the FIB milling process. The red box highlights the region where EDS is performed. **h)** EDS layered map shows three distinct regions in the field of view. The middle plot the row-averaged normalized intensity profile of three representative elements, and the right plot shows the atomic percentage profile of all detected elements. **i)** EDS maps of 4 elements existing in CBST, the atomic percentages are determined to be 62.4% (Te), 21.9% (Sb), 12.0% (Bi), and 3.7% (Cr) respectively.

It is noticeable that at the initial growth of CBST, the lattice constant measured from RHEED pattern shows a quick and nearly linear relaxation from GaAs (111) to CBST, similar to the relaxation behavior of 3D materials [22] despite the fact that the lattice constant relaxation in CBST already finished within 1 QL. The relaxation behavior happens within half a minute which corresponds to 3 atomic layers. The main peak from RHEED also shows no shift through all the growth which an indication of 0 twisted angle. Meanwhile, the EDS have shown a uniform composition distribution through the epitaxial layers, which is also an indication of the coherent strained nature by ruling out the possibility of dominant composition evolution. Further investigations by other ex-situ techniques are also done in this work as discussed below.

*Coherency between epitaxial CBST and GaAs (111) substrate*

We have done X-ray pole figure measurements which show almost 0-degree relative twist angle between CBST and GaAs, despite the existence of a twinning. A STEM (Scanning Transmission Electron Microscope) has also demonstrated a Moiré pattern that corresponds to the 0-degree twisted angle.

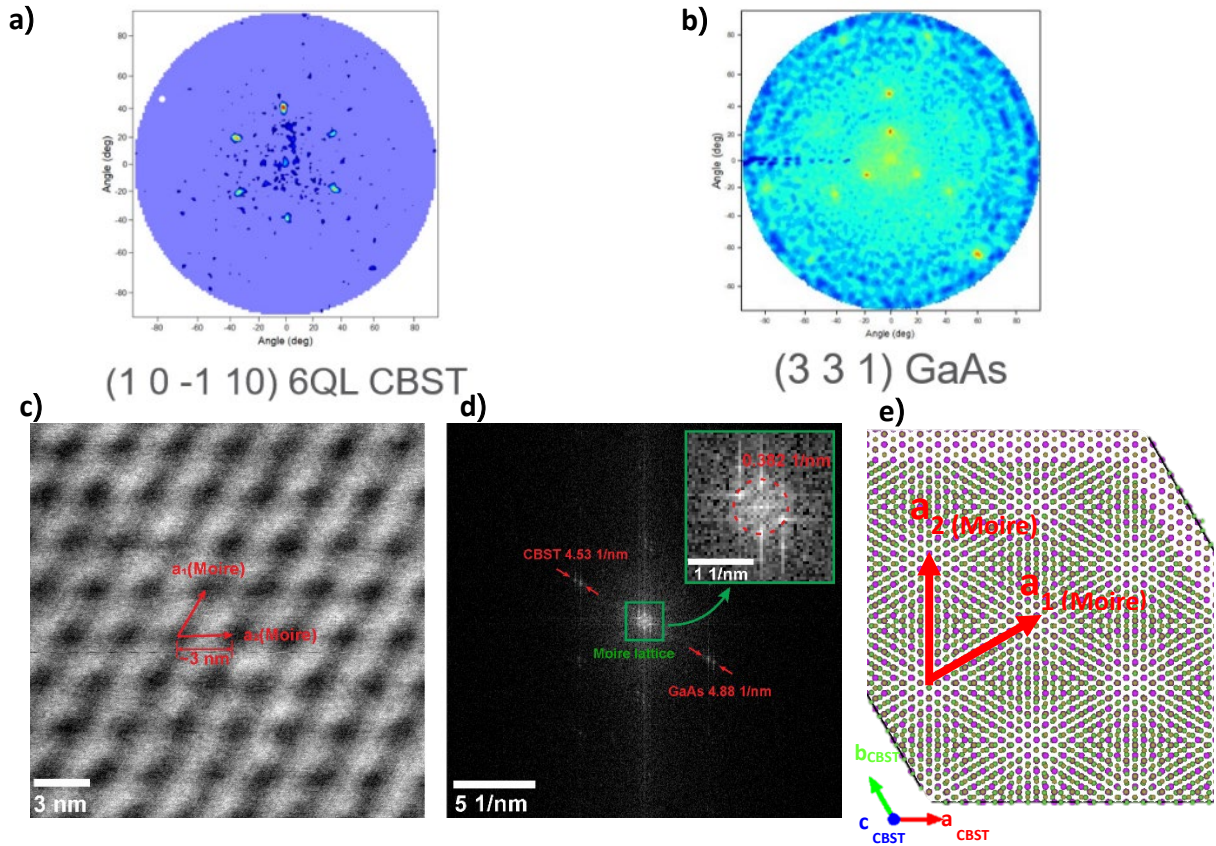

**Fig. 2. Pole Figures from X-ray Reciprocal Space Mapping (RSM) and Moiré pattern from Scanning Transmission Electron Microscope (STEM) show uniform in-plane orientation of CBST on GaAs (111) with no global intrinsic twisted angle observed. a)-b)** Pole figures of CBST (1 0 -1 10) plane (a)) and GaAs (3 3 1) plane (b)). The two scans are made out of the same alignment and show 0 degree twisted angle. It is also noticeable that for each peak on the pole figure of CBST, there is only one dominant peak observed, which is a sign of uniform in-plane orientation of CBST grown on GaAs (111) substrates. Despite the twinning observed, the sample shows good crystallinity quality. **c)** Moiré pattern observed by STEM imaging. The Moiré pattern shows a hexagonal symmetry system, with the local contrast pattern showing regular triangle relationship. The Moiré lattice is measured to be around 2.9-3.0 nm based on the figure. **d)** A FFT of Moiré pattern observed. **e)** A Moiré pattern model （green: substrate or strained epilayer; red/yellow: relaxed epilayer）based on 0 degree twisted angle and the lattices constants of GaAs (111) and CBST (0001). The lattice constant used in the model is 4.35 Å and is calculated based on the composition from EDS measurement Cr: Bi: Sb: Te = 3.7%: 12.0%: 21.9%: 62.4%. Te site from epi-layer and As-corresponding site from substrate layer are set overlapping at the 0-point based on the epitaxial

relationship that Te atom from CBST bottom layer prefer to sit on top of Te atom formed from substitution of As on GaAs substrate. The Moiré pattern shows Moiré lattice parallel to those of the crystal lattice, which is the typical feature of a 0 twisted angle. The Moiré lattice length is 12 times $(2\bar{1}\bar{1})$ spacing of GaAs here, which is about 28 Å, matching the measured Moiré lattice length in c).

Theoretical Moiré lattice constant calculation:
GaAs cubic lattice constant is 5.75 Å, $(2\bar{1}\bar{1})$ spacing is $5.75/\sqrt{2^2 + 1^2 + 1^2} \approx 2.35$ Å.
$2.35 * 12 = 28.2$ Å.

Measured Moiré lattice constant calculation:
k-space first order ring radius is 0.0382 Å^-1. Assuming a perfect triangular Moiré lattice, then the lattice constant is $1/0.0382/(\sqrt{3}/2) \approx 30.2$ Å.

As can be seen in Fig.2., the pole figures show a no twist between CBST and GaAs, with error bar as the only fluctuations here (below 1 °). A Moiré pattern by STEM has further demonstrated the coherency. Detailed analysis can be found under the figures. The Moiré pattern is observed on a 1 QL CBST sample which is as thick as 6 atomic layers. It is an overall effect of both relaxed CBST lattice constant and strained CBST or GaAs (111) lattice constant.

Since each Moiré lattice has hundreds of atoms here due to the 6 atomic layers, the most dominant contrast is the Moiré lattice itself, with the crystal lattice very faint but still noticeable here. However, the FFT (Fast Fourier Transform) has shown a much clearer signals of both GaAs and CBST. It can also be seen from the FFT pattern that CBST, GaAs and Moiré lattice are all aligned in the same direction, which is a further proof of the coherency. The measured twisted angle from FFT here is 0.14 ° ± 0.08 °, which all comes from the error bar.

### *Existence of Strain in the initially growth of CBST on a GaAs (111) substrate*

Reciprocal Space Mapping (RSM) has been a useful tool for strain analysis. [22] Shown in Supplementary Information is the RSM results from a lab-built X-ray diffraction machine. For vdW materials it is difficult to see strain. Here for the first time, we have demonstrated the strain by RSM at CHESS as shown in Fig.3.

We have done the RSM on both Te-annealed GaAs (111) substrate and the 1 QL, 2 QL and 6 QL CBST samples respectively. The CBST samples all show a consistent result on the strained layer, which is indicated by the faint peak across the GaAs peak. The strained layer shows up at the same in-plane lattice constant of GaAs, which is a clear proof of the existence of strain.

It can be seen that a consistent relaxed CBST layer also shows up at the right side of the GaAs or strained CBST. As the thickness increases, the relaxed layers shows a stronger peak and a clearer thickness fringes, which can be used to distinguish the relaxed and the strained layers within CBST. It is also interesting to note that the sharp streaky peaks of CBST all shows clear fringes despite they are only a few nm s thick, which is a sign that the films are very smooth benefited by the qvdWE mode.

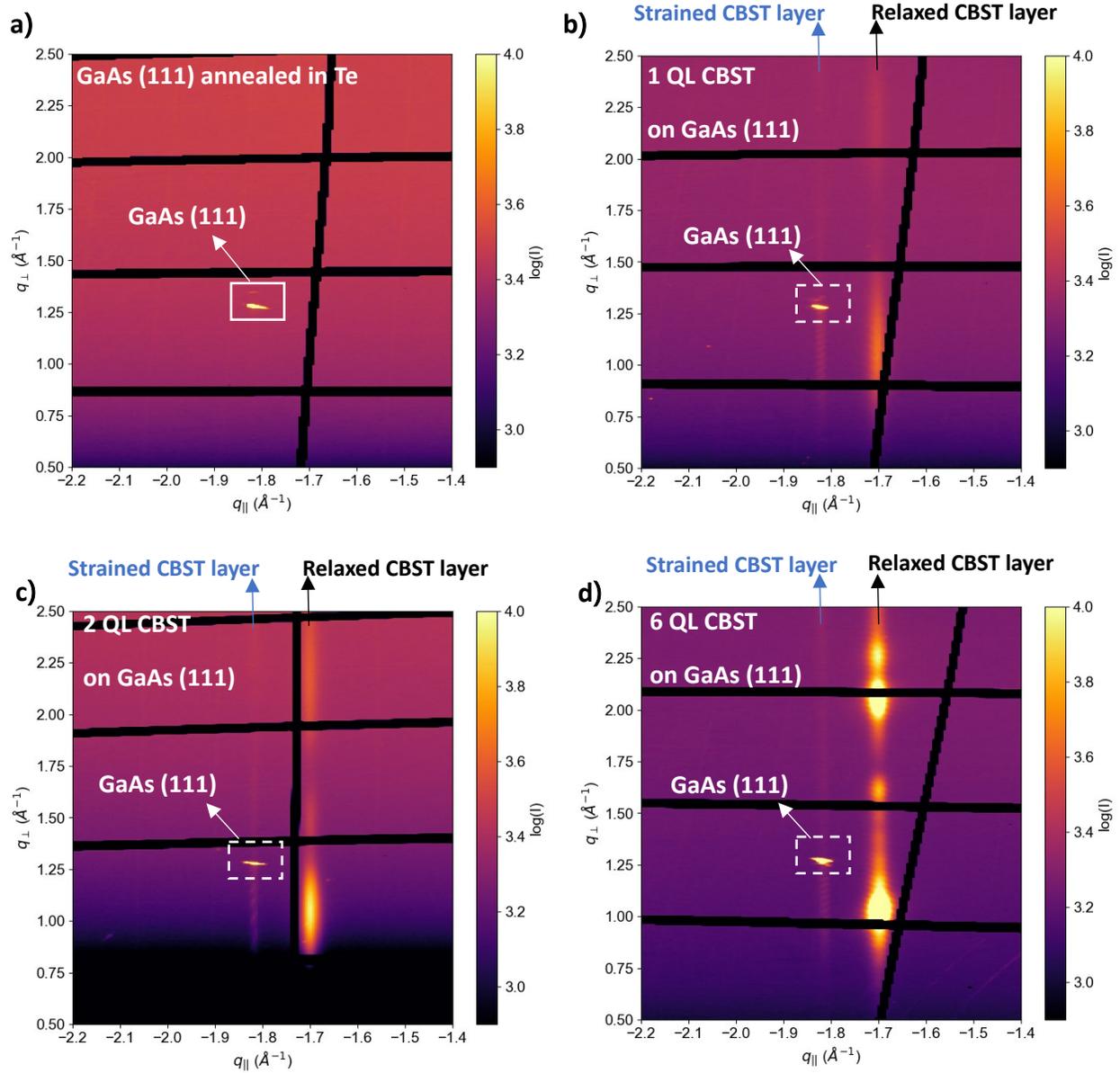

**Fig. 3. Reciprocal Space Mapping (RSM) of a)** GaAs (111) substrate after Te annealing; **b)** 1 QL CBST grown on GaAs (111); **c)** 2 QL CBST grown on GaAs (111); **d)** 6 QL CBST grown on GaAs (111).

The white square marks the peak from GaAs bulk substrates in each sample. The blue arrow shows the peaks which has the same in-plane lattice constant as GaAs in samples only with CBST, indicating the existence of a coherently strained CBST at the initial quasi van der Waals Epitaxy growth mode. The strained layer shows up at a same location (same in-plane lattice constant as GaAs) in samples with different thickness. The relaxed layers are also marked in black arrows in each samples. It can be seen that the relaxed layers show different fringes and intensity due to the difference in their thickness, which is distinguished from the coherently strained layers.

We also also done Raman measurements which shows shifting of peaks and width broadening in Fig. 4. , another sign marching the existence of strain. Peak positions are summarized in Table 1.

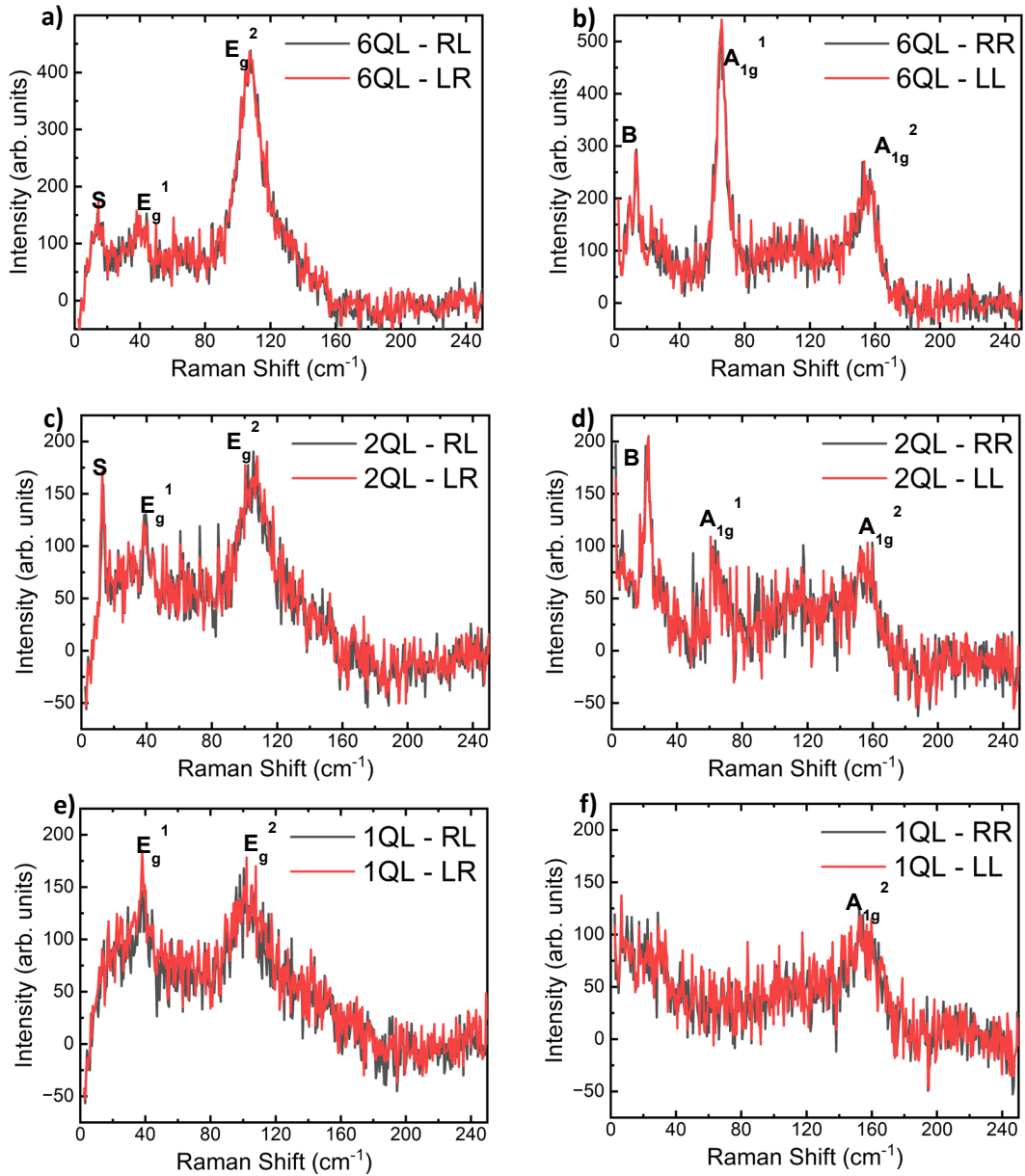

**Fig. 4. Raman Spectroscopy on CBST with 1 QL, 2 QL and 6 QL.**

**a)** Co-circular polarization Raman measurements on 6 QL CBST grown on GaAs (111). **b)** Cross-circular polarization Raman measurements on 6 QL CBST grown on GaAs (111). **c)** Co-circular polarization Raman measurements on 2 QL CBST grown on GaAs (111). **d)** Cross-circular polarization Raman measurements on 6 QL CBST grown on GaAs (111). **e)** Co-circular polarization Raman measurements on 6 QL CBST grown on GaAs (111). **f)** Cross-circular polarization Raman measurements on 6 QL CBST grown on GaAs (111).

(Notations: PP – parallel polarization; CP – cross polarization; LR – co-circular polarization (left circular input and right circular output); RL – co-circular polarization (right circular input and left circular output); LL – cross-circular polarization (left circular input and left circular output); RR – cross-circular polarization (right circular input and right circular output))

|  | CBST thickness | S mode | B mode | $E_g^1$ mode | $E_g^2$ mode | $A_{1g}^1$ mode | $A_{1g}^2$ mode |
|---|---|---|---|---|---|---|---|
| peak position (cm$^{-1}$) | 1 QL | / | / | 36.5±0.8 | 102.0±0.6 | / | 151±1 |
|  | 2 QL | 13±0.1 | 21.8±0.2 | 36±1 | 105.9±0.4 | 64.0±0.6 | 154±1 |
|  | 6 QL | 14.3±0.4 | 13.0±0.3 | 39.4±0.8 | 107.6±0.2 | 65.6±0.1 | 153.9±0.4 |
| peak width (cm$^{-1}$) | 1 QL | / | / | 19±4 | 19±3 | / | 29±4 |
|  | 2 QL | 2.1±0.4 | 5.5±0.6 | 26±6 | 20±2 | 5±2 | 13±4 |
|  | 6 QL | 6±1 | 5.6±0.9 | 13±3 | 17.1±0.7 | 5.6±0.3 | 12±1 |
| peak Intensity (arb. unit) | 1 QL | / | / | 1900 | 2500 | / | 3400 |
|  | 2 QL | 400 | 1300 | 2000 | 3900 | 500 | 900 |
|  | 6 QL | 700 | 1400 | 1500 | 9700 | 4100 | 2700 |

**Table 1. Raman measurement results on CBST of 1 QL, 2 QL and 6 QL grown on GaAs (111) substrates.** There has already been some reports of showing the substrate effect of van der Waals or 2D materials by Raman which use either the $E_g$ mode that is sensitive to strain or the S mode which can be used to demonstrate the in-plane lattice strain specifically. Here we have shown the effect in both modes that corresponds to the in-plane compressive strain.

It is also noticeable that the thinner layers have a broader peak, which is another signature of the overall peak shifting.

With these results, this is the 1$^{st}$ time demonstration of the strain features of qvdWE.

## Discussion

In conclusion, we have shown the coherency by 2 signs: 1) X-ray pole figures and 2) a Moiré pattern by STEM. We have also shown the coherent strain by 3 signs: 1) in-situ RHEED, 2) RSM by a high energy X-ray beam and 3) Raman measurements. These are the first time demonstration of a dominant coherency and strain at the same time in qvdWE modes.

It is important for the effect on the quantization window which is a following up study of this work. This can be used to further demonstrate the materials tuning of the sample quality through growth. The disappear of the strain at grain boundary due to the coherency might be another factor that influencing the composition in quantized samples of MTI besides the doping and substrate band bending effect.

## Methods

### *MBE growth*

The samples are grown by a Perkin-Elmer Molecular Beam Epitaxy system. All the cells used for CBST growth in this work are standard Knudsen effusion cells. Commercial GaAs (111)B substrate is loaded into chamber by Indium mounting at the backside to assist uniform temperature distribution. The substrate is heated up to and kept at 600 °C inside the growth chamber under a high Te pressure ($T_{Te}$ = 380 °C) to remove surface oxide utilizing the surface dissociation of GaAs. The high Te pressure helps reduce the over-annealing of GaAs substrates. Substrates are immediately cooled down to a growth temperature of $T_{sub}$ = 210 °C in Te pressure after the oxide is removed. The samples are grown at the condition of $T_{Cr}$ = 1090 °C, $T_{Bi}$ = 462 °C, $T_{Sb}$ = 300 °C, and $T_{Te}$ = 330 °C. In-situ RHEED is used to monitor the numbers of quintuple layers grown by RHEED intensity oscillation. Typical samples are grown into 6 QL. After growth, samples are cooled down in Te pressure. In this work, we have grown samples at 1 QL, 2 QL and 6 QL respectively using the same growth condition above to demonstrate the strain effect and interfacial properties.

### STEM imaging and EDS analysis

The electron microscopy data is acquired in a FEI Titan 80-300 STEM without an aberration corrector or a monochromator. The energy dispersive spectroscopy (EDS) data is acquired with an Oxford X-MaxN 100TLE 100 mm$^2$ silicon drift detector (SDD). The imaging conditions have an accelerating voltage of 300 kV, 10 mrad convergence semi-angle, and 0.18 nA beam current. The CBST cross section is prepared with gallium focused ion beam (FIB) in a Nova 600 SEM/FIB System. A layer of platinum is deposited with a small FIB current to minimize damage to the top surface. The EDS map has a live time of 10 mins, and the atomic percentages of detected elements are determined with Cliff-Lorimer method.

### X-ray diffraction, Pole Figures and Reciprocal Space Mapping (RSM)

The ω-2θ scan is done by a standard Panalytical powder-compatible X-ray Diffraction machine with a step size of 0.1 ° and a range 10 °– 70 ° which include key peaks from CBST and GaAs.

Pole figures and a RSM at a specific angle are measured by a home-built X-ray diffraction equipment. We used TAD A2 A2 and glancing incidence for the CBST epilayer.

We have also done a 3D RSM at Cornell High Energy Electron Source (CHESS) for CBST grown on GaAs (111) substrates. All of these maps were collected at photon energy of 15 keV, nominal temperature of 100 K from a nitrogen cryostream, collected using the rotation diffraction method. The images were collected with a Pilatus 6M detector at the the QM2 beam line at Cornell High Energy Synchrotron Source.

### Raman Spectroscopy

Raman measurements are done using a home-built cryogenic polarized precision Raman system. The measurements are done with a vacuum level of 1 x 10$^{-7}$ hpa, an excitation laser 532.16 nm (2.33 eV), a laser power 400 μW and a spectrometer window size around 20 nm (around 750 cm$^{-1}$ near 532 nm).

## Acknowledgement

The authors would like to thank Shuyu Cheng, Qing Lou, Xiyue Zhang, Peng Zhang and Ziming Shao for useful discussions.
## Author Contributions

Yuxing Ren grew the CBST samples measured in this work. Lixuan Tai and Xiang Dong have also grown CBST samples from the same batch. Kaicheng Pan and Michael Liao have built the Reciprocal Space Mapping (RSM) model. Kaicheng Pan and Michael Liao have done the pole figures and an RSM measurement. Yueyun Chen has done the STEM and EDS measurements. Benjamin Z. Gregory, Suchismita Sarker, Yuxing Ren and Yifei Sun have done the RSM measurements at Cornell High Energy Synchrotron Source (CHESS). Jin Ho Kang has done the Raman measurements. Noah Bodzin and Kin Wong have done the Focus Ion Beam (FIM) cutting and sample preparations for STEM measurements. Malcolm Jackson, Yuxing Ren and Yueyun Chen have built the model of Moiré lattice. Yuxing Ren have done the analysis work and wrote the manuscript with input from all authors.

## References

1. Alaskar Y, Arafin S, Martinez-Velis I, Wang KL. Heteroepitaxial growth of III–V semiconductors on 2D materials. Two-dimensional Materials-Synthesis, Characterization and Potential Applications. 2016 Aug 31.

2. Kum, H., Lee, D., Kong, W.,…, Kim, J. , Epitaxial growth and layer-transfer techniques for heterogeneous integration of materials for electronic and photonic devices. Nat Electron 2, 439–450 (2019).

3. Wu, Hao, Aitian Chen, Peng Zhang, Haoran He, John Nance, Chenyang Guo, Julian Sasaki et al. "Magnetic memory driven by topological insulators." Nature communications 12, no. 1 (2021): 1-7.2

4. Kou, Xufeng, Murong Lang, Yabin Fan, Ying Jiang, Tianxiao Nie, Jianmin Zhang, Wanjun Jiang et al. "Interplay between different magnetisms in Cr-doped topological insulators." ACS nano 7, no. 10 (2013): 9205-9212.

5. Alaskar, Yazeed, Shamsul Arafin, Darshana Wickramaratne, Mark A. Zurbuchen, Liang He, Jeff McKay, Qiyin Lin, Mark S. Goorsky, Roger K. Lake, and Kang L. Wang. "Towards van der Waals epitaxial growth of GaAs on Si using a graphene buffer layer." Advanced Functional Materials 24, no. 42 (2014): 6629-6638.

6. Ren, Yuxing, Lixuan Tai, Hung-Yu Yang, Xiang Dong, Ting-Hsun Yang, Yaochen Li, and Kang Wang. "Quasi Van der Waals Epitaxy of Magnetic Topological Insulator on GaAs (111) Substrate." In APS March Meeting Abstracts, vol. 2023, pp. Q34-012. 2023.

7. Sever, Vitomir, Nicolas Bernier, Damien Térébénec, Chiara Sabbione, Jessy Paterson, Florian Castioni, Patrick Quéméré et al. "Quantitative STEM HAADF Study of the Structure of pseudo-2D $Sb_2Te_3$ films grown by (quasi) van der Waals Epitaxy." physica status solidi (RRL)–Rapid Research Letters.

8. Tokura, Yoshinori, Kenji Yasuda, and Atsushi Tsukazaki. "Magnetic topological insulators." *Nature Reviews Physics* 1, no. 2 (2019): 126-143.